\documentstyle[amstex,righttag,intlim,twoside,amssymb,epsfig,euscript]{amsart}
\def\mathcal{\cal}
\def\newblock{\hskip .11em plus.33em minus.07em}%
\def\E{\EuScript}
\newcommand{\CinfO}{C^\infty_0}                                 
\newcommand{\Cinf}{C^\infty}                                    
\newcommand{\supp}{\operatorname{supp}}                         
\newcommand{\singsupp}{\operatorname{singsupp}}                 
\newcommand{\WF}{\mbox{\rm WF}}                                 
\newcommand{\cotan}{T^{*}\!}                                    
\newcommand{\Rn}{{{\Bbb R}^n}}                                  
\newcommand{\wick}[1]{:\!{#1}\!:}                               
\newcommand{\wkn}[2]{\wick{\phi^{#1}_{#2}}}                     
\theoremstyle{definition}
\newtheorem{Dfn}{Definition}[section]
\newtheorem{exmp}[Dfn]{Example}
\theoremstyle{plain}
\newtheorem{Lemma}[Dfn]{Lemma}
\newtheorem{Thm}[Dfn]{Theorem}
\newtheorem{Cor}[Dfn]{Corollary}
\newtheorem{Prop}[Dfn]{Proposition}
\theoremstyle{remark}
\newtheorem{rem}{Remark} 
\newtheorem{ack}{Acknowledgement} 
\begin{document}
{\tt \noindent DESY 95-196 \hfill ISSN 0418-9833\\
     gr-qc/9510056 \hfill \\}
\vskip0.7truecm
\author[R.Brunetti, K Fredenhagen and M. K\"ohler]{{\bf
R. Brunetti$^{\rm (1)}$,
K. Fredenhagen$^{\rm (2)}$ and M. K\"ohler$^{\rm (2)}$}\\
\hfill\\
${}^{(1)}$Dip. di Scienze Fisiche\\
           Universit\`{a} di Napoli ``Federico II''\\
           PAD. 19, Mostra d'Oltremare\\
           I--80125 Napoli, Italy\\
\hfill\\
${}^{(2)}$II.\ Institut f\"ur
             Theoretische Physik,\\
             Universit\"at Hamburg\\
             Luruper Chaussee 149\\
             D--22761 Hamburg, Germany}
\thanks{e-mail:\ brunetti@@axpna1.na.infn.it}
\thanks{e-mail:\ i02fre@@dsyibm.desy.de}
\thanks{e-mail:\ mkoehler@@x4u.desy.de}
\title[Microlocal spectrum condition and Wick polynomials]{The microlocal
 spectrum condition and Wick polynomials
 of free fields on curved spacetimes}
\date{\today}
\dedicatory{{\rm PACS numbers: 1110, 0460, 0230}}
\maketitle
\begin{abstract}
  Quantum fields propagating on a curved spacetime are investigated in
  terms of microlocal analysis. We discuss a condition on the wave
  front set for the corresponding n-point distributions, called
  ``microlocal spectrum condition'' ($\mu$SC). On Minkowski space, this
  condition is satisfied as a consequence of the usual spectrum
  condition.  Based on Radzikowski's determination of the wave front
  set of the two-point function of a free scalar field, satisfying
  the Hadamard condition in the Kay and Wald sense, we construct in
  the second part of this paper all Wick polynomials including the
  energy-momentum tensor for this field as operator valued
  distributions on the manifold and prove that they satisfy our
  microlocal spectrum condition.
\end{abstract}
\newpage
\section{Introduction}
\label{sec:intro}
Quantum Field Theory on curved spacetime (QFT on CST) describes
quantum fields propagating under the influence of an external
gravitational field. The main problem which has to be resolved in this
setting is an appropriate formulation of stability.  On Minkowski
space, stability is expressed by the requirement of positive energy,
i.e., the generators of time-like translations are represented as
positive operators on some distinguished Hilbert space.  In the
absence of a time-like Killing vector field a corresponding condition
can not be formulated and there exists no preferred Hilbert space.
These difficulties can best be treated in the algebraic approach to
quantum field theory~\cite{haag:64}. In this approach, the formulation
of a particular model can be divided into two steps. In the first step
an algebra of observables is constructed in terms of commutation or
anticommutation relations. This step was performed by
Dimock~\cite{dimock:80,dimock:82,dimock:92} for the free scalar, Dirac
and electro-magnetic fields on globally hyperbolic spacetimes. In a
second step, a class of states with suitable stability properties has
to be found. States are here considered as expectation functionals,
i.e., normalized positive linear functionals on the algebra of observables.
Via the GNS construction each state induces a
representation of the observable algebra in a Hilbert space with a
cyclic vector but different states may lead to inequivalent
representations.

One approach to characterize the class of states mentioned above is
given by the ``scaling limit criterion'' and the ``principle of local
definiteness'' introduced by Haag, Narnhofer and
Stein~\cite{haagnarnhofer:84} and further investigated by Fredenhagen
and Haag in~\cite{fredenhagen:87}. Both characterizations are designed
for states on general quantum field theoretical models on arbitrary
spacetimes. On the other hand, for linear free models the so-called
Hadamard states are believed to be physical states since the work of
DeWitt and Brehme~\cite{DeWitt:60}. See Kay and Wald~\cite{kay:91} for
a precise definition.  These states are characterized by the short
distance behaviour of their two-point
distributions~\cite{fullingSweenyWald:78}: The singularity structure
of the latter has to satisfy the Hadamard condition, in particular it
is fixed by the underlying geometry.  It was shown in~\cite{Verch:94}
that these states satisfy the condition of local definiteness
and that their local von Neumann
algebras are factors of type $\text{III}_1$ (see
also~\cite{wollenberg:92b}).  Finally, in Hadamard states the
expectation value of the stress energy tensor had successfully been
renormalized~\cite{wald:78}.  Unfortunately the Hadamard condition is
restricted to linear (free) fields.

Radzikowski showed in~\cite{Radzikowski:92} (see
also~\cite{koehler:95}) that the global Hadamard condition can
equivalently be formulated in terms of wave front sets. This
formulation allows the powerful mathematical techniques of microlocal
analysis developed in~\cite{Hoermander:71,Hoermander:72} to be used in
QFT on CST.  Indeed, one of us showed in~\cite{koehler:95} that the
product of two scalar fields on a manifold is a well defined Wightman
field on the GNS Hilbert space of any globally Hadamard product state.
In this work we show a similar result for Wick-powers of scalar fields
on a manifold. Moreover --using wave front sets-- we propose a
microlocal spectrum condition ($\mu{}$SC) for all n-point
distributions of a state, which is a generalization of the usual
spectrum condition to manifolds. The first formulation of such a wave
front set spectrum condition is due to
Radzikowski~\cite{Radzikowski:92}. However, his proposal needed to be
modified, since free fields and products of free fields respectively
provided counterexamples for his definition. It is also shown below
that all Wick-powers of a free scalar field satisfy our $\mu$SC. This
gives, besides the product of two different scalar fields, another
(nontrivial) example for a Wightman field satisfying our new
condition.  We expect the $\mu$SC to be an important ingredient for
the formulation of quantum field theory on curved spacetimes, as well
as a valuable tool for a rigorous formulation of perturbation theory
on manifolds. In addition it might even deepen our understanding of
the Wightman theory on Minkowski space and provide a local
characterization of interaction.

The organization of this work is as follows. After this introduction
some fundamental definitions and results of H\"ormander's microlocal
analysis are restated, i.e., the definition of wave front sets,
theorems on the multiplication of distributions and composition of
distribution valued operators respectively. These mathematical
preliminaries are followed by a section on Hadamard states, in which
Radzikowski's wave front characterization of these states is
presented. In section~\ref{sec:MSC} we propose our new $\mu$SC and
show some fundamental consequences of our definition. Finally in the
last section, Wick polynomials of a free scalar field are constructed.
It is shown that they are (nontrivial) examples for new Wightman
fields on a manifold.

\section{Mathematical Preliminaries}
\label{sec:prim}
This section is mainly written with the purpose to give an as much as
possible self-contained introduction to some definitions and results
of H\"ormander's microlocal analysis, which will be needed in the
sequel of this paper.  Following the definition of the wave front set
of a distribution, we recall H\"ormander's results on the
multiplication of distributions, which extends to the composition of
distribution valued operators. For further details on this
mathematical subject and the proofs of all Theorems stated below in
this section, we refer the interested reader to the monograph of
H\"ormander~\cite{hoermander:analysisI} or to the original literature
(i.e.,~\cite{Hoermander:71,Hoermander:72}).

The theory of wave front sets was developed in the seventies by
H\"ormander together with
Duistermaat~\cite{Hoermander:71,Hoermander:72}, following a work of
Sato~\cite{Sato:69,Sato:70}.The mathematicians use wave front sets
($\WF$) mainly as a tool
in partial differential equations.  Wave front sets are refinements of
the notion of the singular support of a distribution.  One of the main
reasons for using them in favor of the latter is that they provide a
simple characterization for the existence of products of distributions
and eliminate the difference between local and global results.  It is
interesting that Duistermaat and H\"ormander~\cite{Hoermander:72} have
found a link between their microlocal analysis and quantum
field theory, but their results apparently were seldom used in the
physical literature.

The main purpose of microlocal analysis is to shift the study of
singularities from the base space to the cotangent bundle. This is
done by localizing the distribution around the singularity
followed by analysing the result in ``Fourier space''.

Let $u \in {\E D}'(\Rn)$ be a distribution and let
$\phi\in\CinfO(V)$ be a smooth function with support in $V \subset
\Rn$. By a well known argument in distribution theory the
Fourier transform of $\phi u$ yields a smooth function in
frequency space for which the following relation holds:
\begin{equation*}
\widehat{\phi u}(\xi)= <u, e^{-i<\cdot,\xi>}\phi>\, ,
\end{equation*}
where $<\cdot,\cdot>$ denotes dual pairing.
This result implies
\begin{Lemma}\label{Lemma:growestimate}
  Let $u\in{\E D}' (V)$ and let $W$ be an open subset of $V$. Then
  $u|_W \in\Cinf(W)$ if and only if for each $\phi\in\CinfO(W)$ and each
  integer $N\ge 0$ there is a constant $C_{\phi,N}$ such that
  \begin{equation*}
    |<u, e^{-i<\cdot,\xi>}\phi>|\le C_{\phi,N} (1+|\xi|)^{-N}\quad
    \forall\xi\in\Rn \, .
  \end{equation*}
\end{Lemma}
The {\em singular support\/}, $\singsupp u$, of $u\in{\E D}' (V)$ is defined
to be the
complement of the largest open subset of $V$ where $u$ is smooth.
Motivated by the previous Lemma, a refinement of the notion of singular
support which takes into account also the direction in which the
Fourier transform does not strongly decay is the notion of wave front sets.
\begin{Dfn}\label{Dfn:wavefrontset}
  The {\em wave front set}, $\WF(u)$, of $u\in{\E D}' (V)$ is the complement
  in $V\times \Rn\setminus\{0\}$ of the set of points $(x_0 , \xi_0) $
  in $V\times \Rn\setminus\{0\}$ such that for some neighbourhood $U$ of
  $x_0$ and some conic neighbourhood $\Sigma$ of $\xi_0$ we have for each
  $\phi\in\CinfO(U)$ and each integer $N\ge 0$ a constant $C_{\phi,N}$ such
  that
  \begin{equation*}
    |<u, e^{-i<\cdot,\xi>}\phi>|\le C_{\phi,N} (1+|\xi|)^{-N}\quad
    \forall\xi\in\Sigma \, .
  \end{equation*}
\end{Dfn}
Note that a conic set $\Sigma$ is such that if $(x,\xi)\in\Sigma$ then
$(x,t\xi)\in\Sigma$ for all $t>0$.
The following Remarks can be easily proved using Lemma~\ref{Lemma:growestimate}
\begin{rem} \label{remark}\label{page:remark}
$\phantom{c}$\par
  \begin{enumerate}
  \item For $v \in {\E D}'(V)$, $V$ an open subset of $\Rn$,
    with wave front set
    \WF$(v)$, the projection of the wave front set to the base point
    gives the singular support of $v$.
  \item $\WF{}(v)$ is a closed subset of $V\times\Rn \setminus \{0\}$
    since each point $(x,k) \not\in \WF(v)$ has by
    definition an open neighbourhood in $V\times\Rn\setminus \{0\}$
    consisting of such points, too.
  \item\label{rem:wfinklusion} For all smooth test functions $\phi$
    with compact support $\WF(\phi v) \subset \WF(v)$.
  \item For any distribution $v$ with wave front set $\WF(v)$, the
    wave front sets of its partial derivatives are contained in $\WF(v)$.
  \end{enumerate}
\end{rem}
\begin{exmp}
$\phantom{c}$\par
\begin{enumerate}
\item Let $f\in\Cinf(V)\subset {\E D}'(V)$ be a smooth function, then
  $\WF(f)=\emptyset$.
\item Consider the Dirac $\delta$-distribution on ${\Bbb R}^2$. Its easy
  to see that its wave front set is the following
  $
   \WF(\delta(x,y)) = \{ (x,k;y,k') \in {\Bbb R}^2\times{\Bbb R}^{2n}
   \setminus\{0\}| \quad x=y \; ; \; k=-k'\} .
  $
\end{enumerate}
\end{exmp}
 It is worth recalling that the
set of normal coordinates $(x_1,\dots,x_n,\xi_1,\dots,\xi_n)$ of the
cotangent bundle $\cotan\, V$
over the base coordinates $(x_1,\dots,x_n)$ in $V$ allows us
to identify $V\times\Rn$ with $\cotan\, V$ and to consider
$\WF(u)$ as a subset of the cotangent bundle. Since the definition of wave
front sets is local it can be lifted to
manifolds. We give below an intrinsic definition of wave front sets.

Throughout this work a smooth manifold $M$
is a locally euclidean, Hausdorff, second countable topological space
equipped with a smooth structure.
Let $\Omega_{\alpha} (M)$ denotes the complex line bundle of densities
of order $\alpha$ over a smooth manifold $M$, i.e.,
$\Omega_{\alpha} (M) = \cup_{x\in M} \Omega_{\alpha} (T_x M)$
where $T_x M$ denote 
the tangent space at $x$ of $M$ (see~\cite{Duistermaat:FourierIntegral}).
Let then $\CinfO(M,\Omega_{1-\alpha})$ be the vector space
of compactly supported smooth sections of the line bundle $\Omega_{1-\alpha}
\equiv \Omega_{1-\alpha}(M)$ equipped with the usual topology, then,
 ${\E D}'(M,\Omega_\alpha)$, the space of distribution
densities of order $\alpha$ over a smooth manifold $M$, is defined as
the space of
all continuous linear forms on the space $\CinfO(M,\Omega_{1-\alpha})$.
For simplicity we call distribution
densities of order zero {\em distributions} and denote their
corresponding space by ${\E D}'(M)$. It is worth noting that the set
of all smooth functions on $M$ is dense in ${\E  D}'(M)$.
\begin{Dfn}\label{Dfn:reg_dir_points}
  Let $M$ be an n-dimensional smooth manifold with cotangent bundle
  $\cotan M$ and
  take $u\in {\E D}'(M,\Omega_\alpha)$. The point $(x_0, k_0) \in
  \cotan M\setminus\{0\}$ is called a {\em regular directed point\/}
  if and only if for all $s\ge 1$, for all $\lambda_0\in {\Bbb R}^s$
  and for any function $\phi\in C^{\infty}(M\times{\Bbb R}^s,{\Bbb
    R})$ such that $d_x \phi(x_0,\lambda_0)=k_0$ there exists a
  neighbourhood $V$ of $x_0$ in $M$ and a neighbourhood $\Lambda$ of
  $\lambda_0$ in ${\Bbb R}^s$ such that, for all $\rho\in C_0^\infty
  (V,\Omega_{1-\alpha})$ and all $N \ge 0$, we have, uniformly in
  $\lambda\in\Lambda$,
  \begin{equation*}
     |<u,\rho\,\,
    e^{-\imath\tau\phi(\cdot,\lambda)}>|=0(\tau^{-N}) \qquad\mbox{if}
    \quad\tau\longrightarrow \infty\, .
  \end{equation*}
\end{Dfn}
The wave front set, $\WF(u)$, of $u\in{\E D}'(M,\Omega_\alpha)$ is
now the complement in $\cotan M\setminus\{0\}$ of the set of all
regular directed points of $u$. By localization and choice of a
strictly positive density on $M$, such that one can identify
test-densities with compact support with test-functions with compact
support, the definition coincides with that given previously.
Moreover, the same properties as in the Remark on
page~\pageref{page:remark} hold on manifolds.

A useful application of wave front sets is the definition of products
of distributions.  Wave front sets provide a simple characterization
for the existence of such products, which turns out to be
sequentially continuous provided the wave front sets of the
corresponding distributions are contained in a suitable cone in
$\cotan M\setminus\{0\}$.
\begin{Dfn}\label{dfn:seqcontdef}
  Let $\Gamma$ be a closed cone in $\cotan M\setminus\{0\}$ and let
  ${\E D}_{\Gamma}'(M,\Omega_\alpha)$ denote the subspace of
  distributional densities of order $\alpha$ with wave front set
  contained in $\Gamma$. A sequence $\{u_j\}$ of distributional
  densities in ${\E D}_{\Gamma}'(M,\Omega_\alpha)$ converges to a
  distribution $u\in {\E D}_{\Gamma}'(M,\Omega_\alpha)$ iff the following
  conditions hold;
  \begin{enumerate}
   \item $\{u_j \}$ converges to $u$ in ${\E D}'(M,\Omega_\alpha)$ (weakly),
   \item for all $(x_0,k_0)\in (\cotan
     M\setminus\{0\})\setminus\Gamma$, there exists a density of order
     $(1-\alpha)$ $\rho\in\CinfO(M,\Omega_{1-\alpha})$ with
     $\rho(x_0) \neq 0$, a conic neighbourhood $W$ of $k_0$ in
     $\cotan M\setminus\{0\}$ and a function $\phi$ as in
     Definition~\ref{Dfn:reg_dir_points} such that for all $N$,
     \begin{equation*}
       \sup_{\tau\in{\Bbb R}_+}\,\sup_{k\in W}\,\,
       \left [\tau^N |<u-u_j ,\rho\,\,
        e^{-i\tau\phi(\cdot,k)}>|\right ]
        \longrightarrow 0\quad\mbox{if}\quad
        j \longrightarrow\infty .
     \end{equation*}
  \end{enumerate}
\end{Dfn}
\noindent It is worth noting that every subspace
 ${\E D}'_\Gamma(M,\Omega_\alpha)$ contains all smooth densities of order
$\alpha$ with compact support. Moreover, let
$u\in{\E D}_{\Gamma}'(M)$, then there exists a sequence $\{u_j\}$ of
compactly supported smooth functions such that $u_j\to u$ in
${\E D}_{\Gamma}'(M)$. It is therefore possible to choose their supports
in an arbitrary neighbourhood of the support of $u$.
\begin{Thm}\label{Thm:prodDist}
 Let $M$ be a smooth manifold and let $\Gamma$, $\Sigma \in \cotan M
 \setminus\{0\}$ be two closed cones, such that $\Gamma \oplus \Sigma
 := \{(x, k+l) | (x,k) \in \Gamma; (x,l) \in \Sigma\} \subseteq \cotan M
 \setminus\{0\}$. Then the multiplication operator
 \begin{equation*}
   \CinfO(M,\Omega_{\alpha})\times\CinfO(M,\Omega_{\beta}) \ni (u,v)\longmapsto
   u\cdot v \in\CinfO(M,\Omega_{\alpha+\beta})
 \end{equation*}
 extends to a unique sequentially continuous operator from
${\E D}'_\Gamma(M,\Omega_{\alpha})\times
{\E D}'_\Sigma(M,\Omega_{\beta})$
to ${\E D}'_\Lambda(M,\Omega_{\alpha+\beta})$, where
 $\Lambda:=(\Gamma\oplus\Sigma)\cup\Gamma\cup\Sigma$.
 In particular, one finds for the wave front sets:
 \begin{equation}
  \WF(u\cdot v)\subseteq\WF(u)\oplus\WF(v)\cup\WF(u)\cup\WF(v).
 \end{equation}
\end{Thm}
\begin{pf}{}
  In H\"ormander~\cite[Theorem~$2.5.10$]{Hoermander:71} this Theorem
  is proved for the case $\alpha=\beta=0$ only, but the result extends
  to arbitrary values of $\alpha$ and $\beta$: Note first that every
  distribution density of order $\alpha${} can be written as a
  distribution times a smooth positive density of order $\alpha$.
  Using H\"ormander's result the product of these distributions
  multiplied with the tensor product of the corresponding smooth
  densities yields a distributional density of order $\alpha+\beta$,
  which does not depend on the splitting done before.  For the last
  statement recall that two smooth densities differ by a smooth
  function and that the multiplication operator is associative.
\end{pf}{}
\begin{rem}
  Note that for the existence of the product of two distributional
  densities $u$ and $v$ it is sufficient to check that
  $\WF(u)\oplus\WF(v)$ does not contain terms of the form $(x,0)$.
\end{rem}

\subsection{The composition of distribution valued operators}
In this subsection some properties for the composition of two
distribution valued operators are recalled.
Let $M$ and $\tilde{M}$ be two smooth manifolds. Consider an operator $K_1$
from $\CinfO(M,\Omega_1)$ to ${\E D}'(\tilde{M},\tilde{\Omega}_1)$.
By Schwartz nuclear Theorem this operator is
in a one-to-one correspondence to a distribution density
${\cal K}_1\in{\E D}'(M \times \tilde{M},\tilde{\Omega}_1)$, such that,
\begin{equation*}{}
<K_1\rho,\tilde{g}> \, = \, {\cal K_1}(\rho\otimes\tilde{g}) \qquad\forall
\rho\in\CinfO(M,\Omega_1),\,\, \tilde{g}\in\CinfO(\tilde{M}).
\end{equation*}

For the wave front sets of $K_1$ we introduce the following
notations,
\begin{eqnarray*}
 \WF(K_1) & := & \WF({\cal K}_1)\\
 \WF'(K_1) & := & \{(x,h;\tilde{x},\tilde{h})\,\,|\,\,
  (x,h;\tilde{x},-\tilde{h})\in\WF(K_1)\}\\
 \WF_{\tilde M}(K_1) & := & \{(\tilde{x},\tilde{h})\,\,|\,\,
  (x,0;\tilde{x},\tilde{h})\in\WF(K_1)\}.
\end{eqnarray*}

Consider now a second  operator $K_2$ from
$\CinfO(\tilde{M},\tilde{\Omega}_1)$ to ${\E D}'(\Tilde{\Tilde M})$,
where $\Tilde{\Tilde M}$ is a third smooth manifold, whose corresponding
distribution
is denoted by ${\cal K}_2\in{\E D}'(\Tilde{\Tilde M}\times\tilde{M})$.
\begin{Thm}[\mbox{\cite[Theorem~8.2.14]{hoermander:analysisI}}]
  \label{Thm:composition}
  The {\em composition\/} $K_2\circ K_1$ is a well defined operator
  from $\CinfO(M,\Omega_1)${} to ${\E D}'(\Tilde{\Tilde M})$,
  provided that,
  \begin{itemize}
  \item[i)] $K_1$ has proper support, i.e., the inverse image of any
    compact set under the projection $\supp K_1\ni
    ({\Tilde y},x)\mapsto{x}{}$
    is compact,
  \item[ii)] $\WF'_{\tilde M}(K_2)\cap\WF_{\tilde M}(K_1)=\emptyset{}$.
  \end{itemize}
  For the wave front set of $K_2 \circ K_1$ one finds:
  \begin{equation}
    \label{eq:WFcomposition}
    \begin{split}
      \WF{}(K_2 \circ{}K_1) \subseteq & \WF'(K_2) \circ \WF(K_1)\\
      &  \cup \left(
      \WF_{\Tilde{\Tilde{M}}}(K_2) \times M \times \{0\}
              \right)
          \cup \left(
          \Tilde{\Tilde{M}} \times \{0\} \times \WF_M(K_1)
               \right),
    \end{split}
  \end{equation}
  where
  \begin{equation*}
  \begin{split}
  \WF'(K_2) \circ \WF(K_1) =&
  \{(\Tilde{\Tilde{x}},\Tilde{\Tilde{k}};x,k) \in
  (\cotan \Tilde{\Tilde{M}}\times \cotan M)\setminus \{0\} |
  (\Tilde{\Tilde{x}},\Tilde{\Tilde{k}}; \Tilde{x},\Tilde{k}) \in
  \WF'(K_2)\\
  &\qquad\text{and}\quad
  (\Tilde{x},\Tilde{k}; x,k) \in \WF(K_1) \quad
  \text{for some $(\Tilde{x},\Tilde{k}) \in \cotan\Tilde{M}$}\}.
  \end{split}
  \end{equation*}
  Moreover, the composition is sequentially continuous, i.e., if
  $K_{2\epsilon}$ is a sequence of smooth
  $(\WF(K_{2\epsilon})=\emptyset)$ operators with proper supports
  converging\footnote{The topology on the operators is the one
    inherited by that of the corresponding distribution densities.} to
  $K_2$, then the composition $K_{2\epsilon}\circ{}K_1$ converges to
  $K_2\circ{}K_1$, provided the latter composition exists.
\end{Thm}
To shorten the notation we will use the same symbol for the operator
and the distribution density in what follows.
The next Corollary will be useful in section~\ref{sec:MSC}.
\begin{Cor}[\mbox{\cite[Theorem~8.2.13]{hoermander:analysisI}}]
  \label{Cor:partial_smearing}
  Let $K \in {\E D}'({{M}} \times \Tilde{{M}})$ denote an
  operator from $\Cinf_0(\Tilde{M},\tilde\Omega_1)$ to
  ${\E D}'({{M}})$. Then for $\rho\in
  \Cinf_0(\Tilde{M},\tilde\Omega_1)$
   $\WF(K(\cdot,\rho)) \subseteq
  \WF_{\Tilde{M}}(K)$.
\end{Cor}
\section{Hadamard states}
\label{sec:hadamard}
We describe quantum fields propagating on a four dimensional
Lorentz manifold
$(M,g_{ab})$ in terms of the General Theory of Quantized Fields
 (See ~\cite{fredenhagen:91,haag:alg,wald:QFT}).
The manifold is assumed to be globally hyperbolic, i.e., it
admits space-like Cauchy
hypersurfaces. We deal with the G{\aa}rding-Wightman
{}~\cite{WightmanGarding:64} approach to quantum fields and its
algebraic formulation by Borchers and
Uhlmann~(\cite{borchers:62,uhlmann:62}). We refer the reader to the
cited literature for details, but nonetheless let us give some ideas
for completeness. The Borchers-Uhlmann algebra $\cal B$ for the scalar
field mentioned above is the tensor algebra over $\CinfO
(M,\Omega_1)$. A state
$\omega$, defined as a positive linear functional over $\cal B$
consists of a hierarchy of m-point distributions
$\omega=\{\omega_m\}_{m\in{\Bbb N}}$. Every state satisfying local
commutativity fixes --via the following GNS-construction Theorem-- a
Hilbert space, a ``vacuum vector'' and a representation of the algebra
$\cal B$ thus links the algebraic approach to the
Hilbert
space setting of G{\aa}rding and Wightman.
%
\begin{Thm}[GNS-reconstruction]
  For every state $\omega=\{\omega_m\}_{m\in {\Bbb N}}$ on the scalar
  Borchers-Uhlmann algebra there is a GNS-tupel $({\cal
    H}_\omega,{\cal D}_\omega,\phi_\omega,\Omega_\omega)$, unique up to
  unitary equivalence, such that for each $m\geq 1$ and any test
  densities $f_1,\dots,f_m \in \CinfO(M,\Omega_1)$
  \[
  \omega_m(f_1,\dots,f_m) = (\Omega_\omega,
  \phi_\omega(f_1)\cdots\phi_\omega(f_m) \Omega_\omega).
  \]
\end{Thm}
Recall that a GNS-tupel
$({\cal H}_\omega,{\cal D}_\omega,\phi_\omega,\Omega_\omega)$ satisfies the
following properties:
\begin{enumerate}
\item ${\cal H}_\omega$ is a separable Hilbert space, ${\cal D}_\omega$ is a
  dense subspace of ${\cal H}_\omega$ and the GNS-vacuum
  $\Omega_\omega$ is a distinguished vector in ${\cal H}_\omega$.
\item The fields $\phi_\omega$ are operator valued distributions,
  i.e., for all $\Phi,\Psi\in {\cal D}_\omega$ the linear
  mapping
  \[
  (\Psi,\phi_\omega(\cdot)\Phi) \colon
  \CinfO(M,\Omega_1) \ni f
  \mapsto
  (\Psi,\phi_\omega(f)\Phi)
  \]
  is in ${\E D}'(M)$.
\item The subspace ${\cal D}_\omega$ contains $\Omega_\omega$ and is an
  invariant domain for the fields, i.e., for each $f\in
  \CinfO(M,\Omega_1)$ the domain of $\phi_\omega(f)$ contains
  ${\cal D}_\omega$ and $\phi_\omega(f) {\cal D}_\omega \subset
  {\cal D}_\omega$.
\item The fields are hermitian, i.e., for each $f\in
  \CinfO(M,\Omega_1)$, the domain of the adjoint of $\phi_\omega(f)$
  --denoted by $\phi^*_\omega(f)$-- contains ${\cal D}_\omega$ and
  $\phi_\omega^*(f)\supset\phi_\omega(\bar{f})$, where the bar denotes
  complex conjugation.
\item The subspace ${\cal D}_\omega$ is generated by applying finitely many
  smeared field operators to the GNS-vacuum.
\end{enumerate}
If $\phi_\omega$ satisfies the Klein-Gordon equation and
its commutator is given by
\begin{equation*}
  [\phi_\omega(f),\phi_\omega(g)]=E(f\otimes g)
 \quad \forall f,g
  \in \CinfO(M,\Omega_1)
\end{equation*}
where $E$ is the difference between the advanced and retarded
fundamental solution of the Klein-Gordon operator, we call
$\omega$ a state of the {\em Klein-Gordon field \/} over $M$.  It was already
mentioned in the introduction, that not all states $\omega$ are
believed to be physically meaningful. A condition which physically
admissible states should satisfy is the Hadamard
condition~(\cite{DeWitt:60}) which was intensively studied by various
authors (see the references in Fulling's
book~\cite{Fulling:aspects_of_qft}).
A  mathematically precise definition of the
Hadamard condition in terms of boundary values of certain complex
valued functions was given recently by Kay and Wald~\cite{kay:91}.
Radzikowski discovered in his thesis that, equivalently,
Hadamard states can be characterized in terms of their wave front
sets. Using his
results Junker~\cite{junker:95} has been able to construct Hadamard
states for free scalar fields on arbitrary globally hyperbolic
spacetimes.  In this section we recall Radzikowski's wave front set
characterization.
%
\begin{Dfn}
  \label{Dfn:wavefrontscalar}
  Let $\omega$ be a quasifree
  state of the Klein-Gordon field
  over a globally hyperbolic manifold
  $(M,g_{ab})$. Then $\omega$ is a
  Ha\-da\-mard state if and only if its two-point
  distribution $\omega_2$ has wave front set
\begin{equation}\label{eq:wfhadamscalar}
  \WF(\omega_2) = \{ (x_1,k_1),(x_2,-k_2) \in \cotan M^2
  \setminus \{0\} |
  \,\,
  (x_1,k_1) \sim (x_2,k_2); \,\, k_1^0 \geq 0 \}
\end{equation}
where $(x_1,k_1) \sim (x_2,k_2)$ means that there exists a light-like geodesic
$\gamma$ connecting $x_1$ and $x_2$ with cotangent vectors $k_1$ at $x_1$
and $k_2$ at $x_2$.
\end{Dfn}
Recall that a state $\omega$ is called {\em quasifree} iff all its
odd m-point distributions vanish and
\begin{equation}\label{eq:qfprop}
  \omega_m(x_1,\ldots,x_m) = \sum_P \prod_r \omega_2(x_{(r,1)},x_{(r,2)}).
\end{equation}
Here $P$ denotes a partition of the set $\{1,\cdots,m\}$ into
subsets which are pairings of points, labeled by $r$. Note that the
ordering of the points in $\omega_2$ is preserved, e.g.\ $(r,1) <
(r,2)$ and no two arguments are identical. The latter fact ensures the
existence of the product $\prod_r$ whenever $\omega_2(x_i,x_j)$ are
distributions.

For the wave front set of $\omega_m$ one finds
using Theorem~\ref{Thm:prodDist} and Eqn.~(\ref{eq:qfprop})
\begin{equation}
    \label{eq:wf_omega_n}
      \WF(\omega_m)  \subseteq  \left(
                      \bigcup_Q \bigoplus_{r\in Q}
                        \WF\left( \omega_2^r \right)
                    \right),
\end{equation}
where $Q$ denotes a nonempty set of disjoint pairs and where
$\omega_2^r$ is the two-point distribution in the varables
$x_{(r,1)}$, $x_{(r,2)}$ considered as a distribution on $M^n$ and
hence has wave front set
\begin{equation}
    \label{eq:wf_omega_2_r}
    \begin{split}
      &\WF (\omega_2^r )=\{(x_1,0;\dots;x_{(r,1)},k_{(r,1)};
      \dots;x_{(r,2)},k_{(r,2)};\dots;x_n,0)|\\
      &\qquad\qquad
      (x_{(r,1)},k_{(r,1)};x_{(r,2)},k_{(r,2)})\in\WF(\omega_2 )\}\, .
    \end{split}
\end{equation}

Using this characterization one can prove the following
\begin{Cor}[\cite{koehler:95}]\label{Cor:scalar_power}
  Let $\omega^1$ and $\omega^2$ be quasifree Hadamard states for two
  massive Klein-Gordon fields propagating on a globally hyperbolic
  spacetime $(M,g_{ab})$. Then the pointwise product of the
  corresponding m-point distributions exists and gives rise to a new
  Wightman field on this spacetime.
\end{Cor}
\section{Microlocal spectrum condition}
\label{sec:MSC}
In this section we propose a condition on the wave front sets of
states of a quantum field on a smooth manifold $M$, which generalizes
the usual Minkowski space spectrum condition to curved spacetimes.
The idea to use wave front sets for a formulation of some kind of
spectrum condition is due to Radzikowski~\cite{Radzikowski:92}.  For a
motivation recall that the Hadamard condition can be formulated as a
condition on the wave front set of the corresponding two-point
distribution.  Since Eqn.~(\ref{eq:wfhadamscalar}) restricts the
singular support of $\omega_2(x_1,x_2)$ to points $x_1$ and $x_2$
which are null related, $\omega_2$ is smooth for all other points.
The smoothness for {\em space-like\/} related points is known to be
true for quantum field theories on Minkowski space satisfying the true
spectrum condition by the Bargmann-Hall-Wightman Theorem. For
time-like related points however a similar general prediction on the
smoothness does not exist.  In order to include possible singularities
at time-like related points, Radzikowski extended
in~\cite{Radzikowski:92} the right-hand side of
Eqn.~(\ref{eq:wfhadamscalar}) to all causally related points; he
proposed that the wave front set of the two-point distributions of any
physical reasonable state should be contained in this extended set. He
called this proposal the ``wave front set spectrum condition''
(WFSSC).  He also proposed a WFSSC for higher m-point distributions
and linked both to the scaling limit condition
of~\cite{haagnarnhofer:84,fredenhagen:87}: Both definitions imply the
true spectrum condition in the scaling limit if this limit exists
(Theorem~4.11 of~\cite{Radzikowski:92}).  Unfortunately it can be
shown that the m-point distributions for $m > 2$ associated to a
quasifree Hadamard state of a scalar field on a globally hyperbolic
spacetime do not satisfy his WFSSC in general.  Moreover it was shown
in~\cite{koehler:95} that the distributional product of two different
fields gives rise to counterexamples even for his two point WFSSC.
This result is also true for the Wick powers constructed below. Thus,
his original WFSSC needs to be modified.

\begin{figure}[hbtp]
  \begin{center}
    \leavevmode
    \epsfig{file=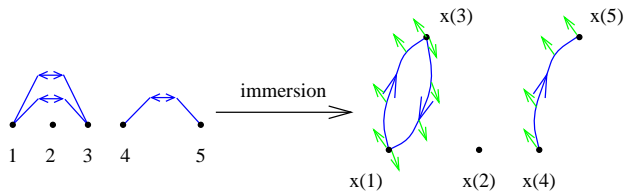}
  \end{center}
  \caption{An immersion of a graph}
  \label{fig:immersion}
\end{figure}

Below we propose a wave front set spectrum condition, which {\em is\/}
satisfied by the examples mentioned above. For a more compact notation
some definitions from graph theory are used: Let ${\mathcal G}_n$
denote the set of all finite graphs with vertices $\{1,\ldots,n\}$,
such that for every element $G \in {\mathcal G}_n$ all edges occur in
both admissible directions. An {\em immersion\/} of a graph $G\in
{\mathcal G}_n$ into some Lorentz manifold $M$ is an assignment of the
vertices of $G$ to points in $M$, $\nu\rightarrow x(\nu)$, and of the
edges of $G$ to piecewise
smooth\label{page:strongmuSC}\footnote{Replacing ``smooth'' by
  ``causal'' or ``light-like'' yields stronger versions of the
  Microlocal Spectrum Condition (Definition~\ref{Dfn:muSC} below).}
curves in $M$, $e\rightarrow \gamma(e)$ with source
$s(\gamma(e))=x(s(e))$ and range $r(\gamma(e))=x(r(e))$ respectively,
together with a covariantly constant causal covector field $k_e$ on
$\gamma$ ($\nabla k_e =0$), such that
\begin{enumerate}
\item If $e$ is an edge from $\nu$ to $\nu'$ then $\gamma(e)$ connects
  $x(\nu)$ and $x(\nu')$,
\item If $e^{-1}$ denotes the edge with opposite direction as $e$,
  then the corresponding curve $\gamma(e^{-1})$ is the inverse of
  $\gamma(e)$,
\item For every edge $e$ from $\nu$ to $\nu'$, $k_e$ is directed
  towards the future whenever $\nu<\nu'$,
\item $k_{e^{-1}}=-k_e$.
\end{enumerate}
(Compare Figure~\ref{fig:immersion})

\begin{figure}[hbtp]
  \begin{center}
    \leavevmode
    \begin{picture}(0,0)%
      \epsfig{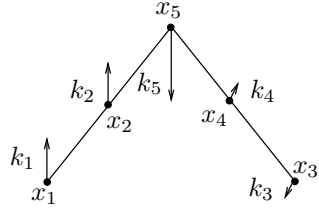}%
    \end{picture}%
    \setlength{\unitlength}{0.00087500in}%
    \begingroup\makeatletter\ifx\SetFigFont\undefined
    \def\x#1#2#3#4#5#6#7\relax{\def\x{#1#2#3#4#5#6}}%
    \expandafter\x\fmtname xxxxxx\relax \def\y{splain}%
    \ifx\x\y   
    \gdef\SetFigFont#1#2#3{%
      \ifnum #1<17\tiny\else \ifnum #1<20\small\else
      \ifnum #1<24\normalsize\else \ifnum #1<29\large\else
      \ifnum #1<34\Large\else \ifnum #1<41\LARGE\else
      \huge\fi\fi\fi\fi\fi\fi
      \csname #3\endcsname}%
    \else
    \gdef\SetFigFont#1#2#3{\begingroup
      \count@#1\relax \ifnum 25<\count@\count@25\fi
      \def\x{\endgroup\@setsize\SetFigFont{#2pt}}%
      \expandafter\x
      \csname \romannumeral\the\count@ pt\expandafter\endcsname
      \csname @\romannumeral\the\count@ pt\endcsname
      \csname #3\endcsname}%
    \fi
    \fi\endgroup
    \begin{picture}(2098,1282)(78,-480)
      \put(631,164){\makebox(0,0)[b]{\smash{\SetFigFont{10}{12.0}{rm}$k_2$}}}
      \put(856,-16){\makebox(0,0)[b]{\smash{\SetFigFont{10}{12.0}{rm}$x_2$}}}
      \put(1036,209){\makebox(0,0)[b]{\smash{\SetFigFont{10}{12.0}{rm}$k_5$}}}
      \put(403,-447){\makebox(0,0)[b]{\smash{\SetFigFont{10}{12.0}{rm}$x_1$}}}
      \put(271,-241){\makebox(0,0)[b]{\smash{\SetFigFont{10}{12.0}{rm}$k_1$}}}
      \put(1706,-426){\makebox(0,0)[b]{\smash{\SetFigFont{10}{12.0}{rm}$k_3$}}}
      \put(1716,164){\makebox(0,0)[b]{\smash{\SetFigFont{10}{12.0}{rm}$k_4$}}}
      \put(1976,-286){\makebox(0,0)[b]{\smash{\SetFigFont{10}{12.0}{rm}$x_3$}}}
      \put(1426, 19){\makebox(0,0)[b]{\smash{\SetFigFont{10}{12.0}{rm}$x_4$}}}
      \put(1151,670){\makebox(0,0)[b]{\smash{\SetFigFont{10}{12.0}{rm}$x_5$}}}
    \end{picture}

  \end{center}
  \caption{A counterexample}
  \label{fig:counterexample}
\end{figure}

We propose the following criterium as a substitute of the usual spectrum
condition for field theories over a globally hyperbolic manifold,
\begin{Dfn}[$\mu$SC]\label{Dfn:muSC}
  A state $\omega$ with m-point distributions $\omega_m$ is said to
  satisfy the {\em Microlocal Spectrum Condition\/} ($\mu{}$SC) if and
  only if, for any $m$,
  \begin{equation}
    \label{eq:muSC}
  \begin{split}
    & \WF(\omega_m)\\
    & \subseteq \biggl\{(x_1,k_1;\ldots;x_m,k_m)\in
  \cotan M^m\setminus\{0\}| \quad \exists G\in {\mathcal G}_m\\
    & \phantom{\subseteq \biggl\{(x_1)}
  \text{~and an immersion $(x,\gamma,k)$ of $G$, such that}\\
    & \phantom{\subseteq \biggl\{(x_1)}
    \quad
    \text{~(i)~}  x_i = x(i) \quad \forall i=1,\ldots, m\\
    & \phantom{\subseteq \biggl\{(x_1)}
    \quad
    \text{(ii)~}  k_i =\sum_{\begin{Sb}
                             e\\
                             s(e)=i
                           \end{Sb}
                           } k_e (x_i)
  \biggr\}=\Gamma_m.\\
\end{split}
\end{equation}
\end{Dfn}
\begin{rem}
  For every set of base points $(x_1,\ldots{},x_m)\in
  \singsupp(\omega_m)$ the first non zero direction $k_l$ in
  the wave front set is future directed.
\end{rem}
\begin{Lemma}\label{Lemma:Gamma_m_additivity}
  The sets $\Gamma_m$ are stable under addition for all $m\in {\Bbb
    N}$, i.e.,
\[
\Gamma_m \oplus\Gamma_m \subseteq \Gamma_m
\]
\end{Lemma}
\begin{pf}
  Let $(${\boldmath $x,k$}$) = (x_1,k_1;\dots,x_m,k_m)$ and
  $(${\boldmath$ x,\tilde{k}$}$) =
  (x_1,\tilde{k}_1;\dots,x_m,\tilde{k}_m)$ be two points in $\Gamma_m$
  with corresponding graphs and immersion $G$, $(x,\gamma,k)$ and
  $\tilde{G}$, $(x,\tilde\gamma,\tilde{k})$ respectively. To prove the
  assertion, it must be shown, that the point $(${\boldmath $x,k_+$}$)
  = (x_1,k_1+\tilde{k}_1;\dots,x_m,k_m+\tilde{k}_m)$ is contained in
  $\Gamma_m$. Let $G_+\in {\cal G}_m$ denote a graph with $m$
  vertices, whose set of edges is the disjoint union of the sets of
  edges
  of $G$ and
  $\tilde{G}$ respectively. The immersions of the latter two graphs
  yield an immersion $(${\boldmath $x,k_+$}$)$ for $G_+$, which
  satisfies i) and ii) of Definition~\ref{Dfn:muSC}. Hence it remains
  to be shown, that $(${\boldmath $x,k_+$}$)$ is contained in
  $\cotan (M^m)\setminus\{0\}$, i.e., that $k_i + \tilde{k}_i \neq 0$ for
  some $i=1,\dots,m$. By definition of $\Gamma_m$, both {\boldmath
  $k$} and {\boldmath $\tilde k$} are nonzero and the first nonvanishing
  entries $k_i$ of  {\boldmath
  $k$} and ${\tilde k}_j$ of  {\boldmath $\tilde k$}
  are elements of ${\overline V}_+ \setminus\{0\}$. Hence, for $l=\min (i,j)$
  we find $k_l + {\tilde k}_l \in {\overline V}_+\setminus\{0\}$.
\end{pf}
To show that there exist non trivial states which satisfy our
Definition we prove the following
\begin{Prop}\label{Prop:qfHadam_muSC}
  Let $\omega$ denote a quasifree Hadamard state for the Klein-Gordon
  field on a globally hyperbolic manifold $(M,g_{ab})$, then $\omega$
  satisfies the $\mu$SC.
\end{Prop}
\begin{pf}
  Note first that all odd m-point distributions vanish by assumption,
  hence $\omega_m$ satisfies the {$\mu$SC} trivially for odd m.
  Consider now the two-point distribution $\omega_2$. Its wave front
  set is given explicitly by Eqn.~(\ref{eq:wfhadamscalar}) and
  obviously satisfies the {$\mu$SC}.  For a general m-point
  distribution consider the representation given by
  Eqn.~(\ref{eq:qfprop}), which states that $\omega_m$ is a sum of
  tensor products of two-point distributions. Hence, there exists a
  disconnected graph $G_m\in {\cal G}_m$ together with an immersion
  $(x,\gamma,k)$, such that (i) and (ii) of Definition~\ref{Dfn:muSC}
  are satisfied, i.e., $G_m$ consists of subgraphs $G_2\in{\cal G}_2$,
  such that the immersion $(x,\gamma,k)$ restricted to these subgraphs
  is compatible with the wave front set of the corresponding two-point
  distribution.
\end{pf}
\begin{rem}
  $\phantom{c}$\par
  \begin{itemize}
  \item A complete analogue of this Proposition should be valid in the
    case of the Dirac equation, since Hadamard states for the latter
    are obtainable by applying the adjoint of the Dirac operator
    to a suitable (auxiliary) Hadamard state of the `squared' Dirac
    equation. For fixed spinor indices the wave
    front set of the latter is contained in the r.h.s. of
    Eqn.~(\ref{eq:wfhadamscalar}) and derivatives do not enlarge the
    wave front set.
  \item Junker~\cite{junker:95} rigorously proved recently that the
    adiabatic vacuum states on a Robertson-Walker spacetime are
    globally Hadamard states. Hence they satisfy the $\mu$SC, too.
  \end{itemize}
\end{rem}
States satisfying the $\mu$SC obey some nice properties outlined in
the following Theorems.
\begin{Thm}\label{Thm:prod_n-pointDist}
  Let $\omega^1$ and $\omega^2$ be two states satisfying the $\mu$SC.
  Then the pointwise products of their corresponding n-point
  distributions exist and define a new state satisfying the $\mu$SC.
\end{Thm}
\begin{pf}
  To prove the existence of the product, it is --by
  Theorem~\ref{Thm:prodDist}-- sufficient to show that the sums of
  $\WF(\omega^i_m)$ $(i=1,2)$ do not intersect with the zero section.
  Now by assumption $\WF(\omega^1_m)$ and $\WF(\omega^2_m)$ are both
  contained in the set $\Gamma_m$, which --by
  Lemma~\ref{Lemma:Gamma_m_additivity}-- is stable under addition.
  Hence $\WF(\omega^1_m)\oplus\WF(\omega^2_m) \subseteq \Gamma_m \subset
  \cotan M^m \setminus\{0\}$, which ensures the existence of the
  products and also implies that they satisfy the $\mu$SC.

  In order to prove that these new m-point distributions yield a
  state, i.e., satisfy Wightman positivity, we consider the {\em
    tensor\/} product of $\omega^1$ and $\omega^2$. This is a state on
  the Borchers-Uhlmann algebra of two commuting scalar
  fields. Positivity for this state implies for every test densities
  $f_j\in \CinfO(M^j,\Omega^j_1)$, $g\in\CinfO(M^2,\Omega^2_1)$,
  \begin{equation}
    \label{eq:postensorprod}
    \begin{split}
      0 \leq &
      \sum_{m,n} \int
      (\omega_1)_{n+m} (x_n,\dots,x_1,x'_1,\dots,x'_m)
      (\omega_2)_{n+m} (y_n,\dots,y_1,y'_1,\dots,y'_m)\\
      & \phantom{      \sum_{m,n} \int}
      \times\,\,\overline{f_n(x_1,\dots,x_n)} {f_m(x'_1,\dots,x'_m)}
      \prod_{i=1}^n \overline{g(x_i,y_i)} \prod_{i=1}^m {g(x'_i,y'_i)}
    \end{split}
  \end{equation}
  Now choose a family of real test densities
  $\{g_\epsilon\in\CinfO(M^2,\Omega^2_1)\}_{0<\epsilon\leq 1}$, such
  that the limit $\epsilon\rightarrow 0$ equals the Dirac
  $\delta$-distribution. Inserting these densities into
  Eqn.~(\ref{eq:postensorprod}), the limit $\epsilon\rightarrow 0$
  exists by the consideration above and yields
  \begin{equation}
    \label{eq:limitprodomega2}
    \begin{split}
      & \sum_{m,n} \int
      \left( (\omega_1)_{n+m}
      (\omega_2)_{n+m}  \right) (x_n,\dots,x_1,x'_1,\dots,x'_m)\\
      & \phantom{\sum_{m,n} \int}\times\quad{}
      \overline{f_n(x_1,\dots,x_n)}
      {f_m(x'_1,\dots,x'_m)}.
    \end{split}
  \end{equation}
  Using the sequential continuity due to Theorem~\ref{Thm:prodDist} we
  conclude that~(\ref{eq:limitprodomega2}) is greater or equal to
  zero, which is the desired positivity. This finishes the proof.
\end{pf}
Let $\omega$ denotes a state on the Borchers-Uhlmann algebra ${\cal B}$
and let $({\cal H}_\omega, {\cal D}_\omega,\phi_\omega,\Omega_\omega)$ be the
associated GNS-tupel. The {\em folium\/} of $\omega$ consists of
finite convex linear combinations $\tilde\omega$ of states induced by
vectors in ${\cal D}_\omega$:
\begin{equation}\label{eq:omegafolium}
  \tilde{\omega}(A) = \text{Tr} ( \rho \pi_\omega(A)) \qquad \forall
  A \in {\cal B},
\end{equation}
where $\text{Tr}$ denotes the trace in ${\cal H}_\omega$,
$\pi_\omega$ is the representation of ${\cal B}$ associated to
$\omega$ and
\[
\rho= \sum_{i=1}^N |\Psi_i><\Psi_i| \qquad \Psi_i \in {\cal D}_\omega,
\] is some density matrix.
Contrary to the usual spectrum condition, the $\mu$SC does not
characterize a distinguished state. Instead it is a property of a
whole folium:
\begin{Thm}
  Let $\omega$ be a state satisfying the $\mu$SC. Then the $\mu$SC is
  satisfied for all states in the folium of $\omega$.
\end{Thm}
\begin{pf}
  Consider a state $\tilde\omega$ in the folium of $\omega$.  All
  m-point distributions $\tilde{\omega}_m$ of $\tilde{\omega}$ are
  finite linear combinations of $l+m$-point distributions of $\omega$
  smeared with suitable test functions from both ends, i.e.,
  \[
   \tilde{\omega}_m(x_1,\cdots,x_m) = \sum_{l}
   \omega_{l+m}
      (f_{j_1},\cdots, f_{j_k},x_1,\cdots,x_m,f_{j_{k+1}},\cdots,f_{j_l}).
  \]
  It is therefore sufficient to show that for all $m\in {\Bbb N}$
  \begin{equation}
    \label{eq:WFomega_n_smeared}
  \Gamma :=\WF(\omega_{l+m}
      (f_{j_1},\cdots, f_{j_k},x_1,\cdots,x_m,f_{j_{k+1}},\cdots,f_{j_l})
     ) \subseteq \Gamma_m,
  \end{equation}
  where $\Gamma_m$ was defined in Definition~\ref{Dfn:muSC} above.

  Using Corollary~\ref{Cor:partial_smearing} we obtain for the
  l.h.s.\ of~(\ref{eq:WFomega_n_smeared})
  \begin{equation}
    \label{eq:WFomega_n_smeared_pf}
    \begin{split}
    \Gamma&\subseteq\{(x_1,k_1;\cdots;x_m,k_m)\in \cotan M^m\setminus\{0\}|\\
     &
     (y_1,0;\cdots;y_{k},0;x_1,k_1;\cdots;x_m,k_m;y_{k+1},0;\cdots;y_l,0)
     \in \WF(\omega_{l+m}) \subset \Gamma_{l+m} \}.
   \end{split}
  \end{equation}
  Moreover, since $\omega_{l+m}$ satisfies the $\mu$SC by assumption,
  we have for every $(y_1,0;$ $\cdots$ $;x_i,k_i;$ $\cdots$ $;y_l,0) \in
  \WF(\omega_{l+m})$ a corresponding graph $G_{l+m} \in {\cal
    G}_{l+m}$ together with an immersion $(x,\gamma,k)$, such that the
  covector fields $k_e$ are zero whenever $\gamma(e)$ does {\em not\/}
  connect two points in $\{x_1,\cdots,x_m\}$. For the last statement
  note first that the direction associated to $y_1$ vanishes by
  (\ref{eq:WFomega_n_smeared_pf}). Moreover, all causal covector
  fields associated to curves $\gamma$ {\em starting\/} at $y_1$ are
  directed towards the future by the definition of an immersion, hence
  using property (ii) of (\ref{eq:muSC}), $k_e=0$ whenever $\gamma$
  starts {\em or ends\/} at $y_1$.

  Consider now the point $y_2$. The direction associated to this point
  is again zero by assumption. Using the properties of the immersion
  and the previous result for covector fields along curves {\em
    ending\/} at $y_1$, one finds that the covector fields $k_e$ for
  all curves {\em starting\/} at $y_2$ are either future directed or
  zero.  As in the previous case this implies $k_e=0$ for {\em all\/}
  curves starting or ending at $y_2$.  By induction, this result
  extends up to $y_{k}$ and analogously from $y_{l}$ down to
  $y_{k+1}$. We may therefore remove from the graph $G_{l+m}$ all
  points $y_1,\cdots,y_{k}$, $y_{k+1},\cdots,y_l$ together with all
  lines starting or ending at these points. The result is a graph
  $G_m\in {\cal G}_m$ together with an immersion $(x,\gamma,k)$ such
  that (i) and (ii) of (\ref{eq:muSC}) are satisfied. This
  completes the proof.
\end{pf}
The following Theorem shows that our new spectrum condition is
compatible with the usual Minkowski space spectrum condition.
\begin{Thm}\label{Thm:SCimpliesmuSC}
  Let $\omega$ be a state for a quantum field theoretical model on
  Minkowski space, whose m-point distributions $\omega_m$ satisfy the
  Wightman axioms. Then $\omega$ satisfies the $\mu$SC.
\end{Thm}
\begin{pf}
  By assumption all m-point distributions $\omega_m$ satisfy the usual
  spectrum condition. Hence for every $m\in{\Bbb N}$  the
  support of the Fourier-transform of $\omega_m$ is contained in the
  following set:
  \[
  \supp(\hat{\omega}_m) \subseteq \{(k_1,\cdots,k_m) \in
  {\Bbb R}^{4m}  | k_1,k_1+k_2,\cdots, \sum_{i=1}^{m-1} k_i
  \in {\overline V}_+, \sum_i k_i = 0\}=: \Gamma_0 \, .
  \]

   For the corresponding wave front set we obtain by
  Proposition~8.1.7 of~\cite{hoermander:analysisI}:
  \[
  \WF(\omega_m) \subseteq {\Bbb R}^{4m} \times \Gamma_0 \setminus\{0\}
  \equiv {\Bbb R}^{4m} \times \Gamma.
  \]
  It is therefore sufficient to show that ${\Bbb R}^{4m} \times \Gamma$ is
  contained in the r.h.s.\ of (\ref{eq:muSC}).

  Let $(x_1,k_1;\cdots;x_m,k_m)$ be an arbitrary element in ${\Bbb
    R}^{4m} \times \Gamma$. Let $G\in {\cal G}_m $ be a graph such that every
  vertex $i$ is connected with its next-neighbour only, i.e., $G$ is a
  simple chain. An immersion $(x,\gamma,k)$ of this graph into
  Minkowski space is given by:
  \begin{enumerate}
  \item Assign to every vertex $i$ the point $x(i):=x_i$.
  \item The curves $\gamma_{i,j}$ are  straight lines connecting $x_i$
    and $x_j$ $(j=(i-1),(i+1))$.
  \item The vector fields $k_{i,i+1}$ are chosen  such that $k_{i,i+1} =
    \sum_{j=1}^{i} k_j$ for all $1\leq i \leq m-1$.
  \end{enumerate}
  Note that all vectors $k_{i,i+1}$ are in the closed forward
  light-cone by assumption and $k_i \equiv k_{i-1,i}-k_{i,i+1}$.
  Obviously this choice is compatible with the $\mu$SC, which shows
  that $(x_1,k_1;\cdots;x_m,k_m)$ is contained in the r.h.s.\ of
  (\ref{eq:muSC}).
\end{pf}
It is worth noting, that Theorem~\ref{Thm:SCimpliesmuSC} seems to be
wrong if
we replace the $\mu$SC by one of its stronger versions as suggested in
the footnote on page~\pageref{page:strongmuSC}: Consider a linear functional
satisfying the usual spectrum condition together with the property
that some point $(x_1,k_1;\cdots; x_5,k_5) \in \cotan M^5\setminus\{0\}$ with
\begin{enumerate}
\item $(x_1,k_1) \sim (x_2,k_2)$ and $k_1 \in {\overline V}_+$,
\item $(x_3,k_3) \sim (x_4,k_4)$ and $k_3 \in {\overline V}_-$ and $k_1
  +k_3 \in {\overline V}_+$
\item $k_1+k_2 = -k_5$,
\end{enumerate}
is in the wavefront set of its corresponding 5-point distribution. (Compare
Figure~\ref{fig:counterexample}).
For such a 5-point distribution there exists no graph $G\in {\cal
  G}_5$ together with an immersion, such that the edges are assigned
to light-like or causal curves {\em and\/} such that the conditions in
Definition~\ref{Dfn:muSC} are satisfied with this immersion. The last
statement remains true even if we use the physical requirement that
this functional satisfies locality. It is not obvious whether {\em
  positive\/} linear functionals with this properties, i.e., states,
exist. The latter would provide a counterexample to the stronger
version of Theorem~\ref{Thm:SCimpliesmuSC} mentioned above.  On the
other hand all states for {\em free\/} field models --including the
Wick powers defined below--, which arise from Hadamard states, satisfy
the stronger versions of our $\mu$SC. One might therefore conjecture,
that the fulfillment of the latter is {\em characteristic\/} for free
field models.
\section{Definition of Wick polynomials for free fields on a manifold}
\label{sec:definition}
Let $\omega$ be a quasifree state of the Klein-Gordon field
over a globally hyperbolic spacetime
$(M,g_{ab})$ satisfying the $\mu$SC.  We define a normal ordering
prescription with respect to $\omega$ as on Minkowski space:
\begin{Dfn}\label{Dfn:normalorder}
  Let $({\cal H}_\omega, {\cal D}_\omega,\phi,\Omega_\omega)$ be the GNS-tupel
  associated to $\omega$. A {\em normal ordering prescription\/} $:~:$ for the
  operator valued distributions $\phi$ is defined by the following
  recursion relation:
  \begin{eqnarray*}
    \wkn{}{} & \;=\; &\phi \\ \wick{\phi(x_1) \cdots \phi(x_{n+1})} &
    \;=\; & \wick{\phi(x_1) \cdots \phi(x_n)} \phi(x_{n+1}) \\ & & -
    \sum_l \wick{\phi(x_1) \cdots\Hat{\phi}(x_l)\cdots \phi(x_n)}
    \omega_2(x_l,x_{n+1}),
  \end{eqnarray*}
  where $\Hat{~}$ means omitting the corresponding element and
  $\omega_2$ denotes the two-point distribution of the state $\omega$.
\end{Dfn}
The normal product
\begin{equation}
  \label{eq:normalorder}
  \wick{\phi(x_1) \cdots \phi(x_n)} \,\equiv\,
  \wkn{\otimes n}{}(x_1,\cdots,x_n)
\end{equation}
is a well defined operator valued distribution on ${\cal H}_\omega$,
since the fields are smeared individually. It is symmetric in its
arguments.  For the subsequent definition of Wick powers, i.e., for
the restriction of Eqn.~(\ref{eq:normalorder}) to the diagonal, some
auxiliary operators are needed:
\subsection{Auxiliary Wick monomials}
\label{sec:auxWick}
Let $\Delta\in {\E D}'(M^n\times M)$ be a distribution, such that for
all smooth test densities $f=f(x_1,\cdots,x_n;x)d\mu_1\cdots d\mu_n
d\mu \in\Cinf_0(M^n \times M,\Omega^{n+1})$
\[
\Delta(f) \equiv \int f(x,\cdots,x;x) d\mu.
\]
It is obvious that $\Delta$ is properly supported. For its wave front set
one finds --using the fact that on Minkowski space $\Delta$ is a
product of Dirac $\delta$ distributions--
\begin{equation*}
  \WF(\Delta) = \{(x,k_1;\cdots;x,k_n;x,k) \in
  \cotan M^{n+1}\setminus\{0\} | \sum k_i = - k \}
\end{equation*}
\begin{equation*}
  \WF_{M^n}(\Delta) = \{(x,k_1;\cdots;x,k_n) \in \cotan M^n \setminus
  \{0\} | \sum k_i = 0 \}
\end{equation*}
For the rest of this paper $\{\Delta_\epsilon\}_{0<\epsilon<1}$ will denote
a family of smooth operators from $\Cinf_0(M,\Omega_1)$ to
${\E D}'(M^n)$ converging to $\Delta$ in ${\E D}'(M^n\times M)$ if
$\epsilon \rightarrow 0$. Without loss of generality we will further assume
that all operators $\Delta_\epsilon$ are real.
\begin{Dfn}\label{Dfn:auxwick}
Given the normal ordering prescription and the family
$\{\Delta_\epsilon\}$ of smooth operators we define the
{\em auxiliary Wick monomials} to be the following operator valued distribution
\begin{equation}\label{eq:aux_wick_monomial}
\wick{\phi_\epsilon^n}\! (f)
  \;:=\; \left( \wkn{\otimes n}{} \circ\, \Delta_\epsilon \right)\!
 (f) \qquad \forall f\in \CinfO(M,\Omega_1),
\end{equation}
where $\circ$ denotes composition of operators in the sense of
Theorem~\ref{Thm:composition}.
\end{Dfn}
Then the {\em Wick monomial\/} $\wkn{n}{}$
should be an operator valued distribution on ${\cal H}_\omega$ formally
defined as the limit:
\begin{equation}\label{eq:WickDfn}
  \wkn{n}{} (f)
  \;=\; \lim_{\epsilon \rightarrow 0} \wkn{n}{\epsilon}\! (f)
  \;=\;  \lim_{\epsilon\rightarrow 0} (\wkn{\otimes n}{}
     \circ\, \Delta_\epsilon) (f) \qquad \forall f\in
  \CinfO(M,\Omega_1),
\end{equation}
It is one of the main results of this paper (See
Theorem~\ref{Thm:WickWightman} below), to show that
Eqn.~(\ref{eq:WickDfn}) above can be defined rigorously and yields new
Wightman fields on the manifold. To achieve this goal let us first
investigate the ``vacuum'' expectation values for products of auxiliary
Wick monomials, i.e., their m-point distributions.
\begin{Prop}\label{Prop:limitwickaux}
  For the m-point distributions of products of auxiliary Wick
  monomials the limit $\epsilon\rightarrow 0$ exists in the sense of
  sequentially continuous convergence for distributions. For
  the corresponding kernels one finds:
  \begin{equation}
   \label{eq:prodauxwickt}
    \begin{split}
     &{\cal W}^{(n_1,\cdots,n_m)}_m (x_1,\cdots,x_m)\\
     &  :=\; \lim_{\epsilon_1\cdots \epsilon_m \rightarrow 0}
        (\Omega,\wkn{n_1}{{\epsilon_1}}(x_1)\cdots
      \wkn{n_m}{{\epsilon_m}}(x_m)\Omega)\\
     &  = \sum\begin{Sb}
       a_{i,j}\\ 1\le i<j\le m
       \end{Sb}
       \left[ \prod_{1\le i<j\le m}\,
       \dfrac{\omega_2\left(x_i, x_j\right)^{a_{i,j}}}{a_{i,j}!}\right ]
       n_1 !\, n_2 !\cdots
       n_m !
    \end{split}
   \end{equation}
   where $a_{i,j}$ is the matrix containing the number of pairings
   between points $x_i$ with $x_j$ $(i,j=1,\dots,m)$ and the sum is taken
   over all possible choices of the $a_{i,j}$ such that $\sum_i
   a_{i,j}=n_j$ for all $j=1,\ldots,m$.
\end{Prop}
\begin{pf}
  We refrain from giving full details of the proof since the
  combinatorial arguments are well-known (see,
  e.g.,~\cite{hepp:renormalisation}).  The only crucial point is to
  show that the r.h.s\ of Eqn.~(\ref{eq:prodauxwickt}) which contains
  products of two-point distributions is well defined, since --by
  Theorem~\ref{Thm:prodDist}-- we have sequential continuity
  whenever the limit exists. On the other hand the existence of the
  limit is an immediate consequence of
  Theorem~\ref{Thm:prod_n-pointDist}, which applies, since all
  quasifree Hadamard states satisfy the $\mu$SC.
\end{pf}
We will show below that the m-point distributions
${\cal W}^{(n_1,\cdots,n_m)}_m $ for all $n_i \leq N$, $1\leq i \leq m$
satisfy positivity and local commutativity, i.e., they are Wightman
distributions. Now every hierarchy of Wightman distributions gives
rise to a {\em state} on the corresponding Borchers-Uhlmann algebra
satisfying local commutativity. Hence we may apply the GNS-reconstruction
Theorem to obtain a GNS-tupel
$({\cal H}_N, {\cal D}_N,\wkn{1}{},\dots,\wkn{N}{},\Omega_N)$, which contains
the
Wick monomials $\wkn{1}{}$, $\dots$, $\wkn{N}{}$ as Wightman fields on
${\cal H}_N$ with dense invariant domain ${\cal D}_N$.  Since $\wkn{}{}
\equiv \phi$, one obviously has ${\cal H}_\omega \subseteq {\cal H}_N$
and ${\cal D}_\omega \subseteq {\cal D}_N$, where
${\cal H}_\omega$ and ${\cal D}_\omega$
are the GNS-Hilbert space and dense invariant domain for the original
field $\phi$ in the state $\omega$ respectively. Now ${\cal D}_\omega$ is
already dense in ${\cal H}_N$, since for every $\Psi\in {\cal D}_N$ there
exists
a sequence of vectors in ${\cal D}_\omega$ converging to $\Psi$. To obtain
the latter replace all Wick monomials in the expression for $\Psi$ by
their corresponding auxiliary Wick monomials. The convergence of the
sequence obtained in that way is an immediate consequence of the
continuity property of the corresponding m-point distribution stated
in Proposition~\ref{Prop:limitwickaux} above. We conclude that
${\cal H}_N$ and ${\cal H}_\omega$ and $\Omega_\omega$ and $\Omega_N$
respectively are equal, which means that the Wick monomials can be
represented as Wightman fields in the Hilbert space ${\cal H}_\omega$.
The subspace ${\cal D}_\omega$ may {\em not\/} be an invariant
domain for the Wick monomials in general, but due to the following
Theorem, it is always a core for the closure of
the Wick monomials with respect to the ``universal'' domain ${\cal D} \equiv
\cup {\cal D}_N$. This result ensures that we ``do not loose information''
by restricting the Wick monomials to the subspace
${\cal D}_\omega$.
\begin{Thm}
  The domain ${\cal D}_\omega$ is a core for the closure of any Wick monomial
  $\wkn{n}{}$ with respect to ${\cal D}$.
\end{Thm}
\begin{pf}
  To prove the Theorem, we have to show that for all test
  densities $f\in \CinfO(M,\Omega_1)$ the closure of $\wkn{n}{}\! (f)$
  restricted to ${\cal D}_\omega$
  coincides with the closure of $\wkn{n}{}\! (f)$
  defined on ${\cal D}$. We note first that all Wick monomials are hermitian,
  hence their closure with respect to ${\cal D}$ exist. Consider an arbitrary
  element $\mbox{$\{\wkn{n}{}\! (f) \Psi,\Psi\}$} \in
  \overline{\Gamma(\wkn{n}{}\! (f))}$ in the closure of the
  graph of $\wkn{n}{}\! (f)$, where $f$ is a test density as above. We will
  show below that there exists a sequence $\tilde{\Psi}_j \in {\cal D}_\omega$
  converging to $\Psi\in {\cal D}$, such that $\wkn{n}{}\! (f) \tilde{\Psi}_j$
  converges to $\wkn{n}{}\! (f)\Psi$.

  Let $\{\Psi_j\}_{j\in {\Bbb N}}$ be a sequence of vectors in ${\cal D}$
  converging to $\Psi$, such that $\wkn{n}{}\! (f) \Psi_j$ converges to
  $\wkn{n}{}\! (f) \Psi$. Our assumption guarantees the
  existence of such a sequence and that every $\Psi_j$ is a polynomial
  of Wick monomials applied to the GNS-vacuum. Replacing every Wick
  monomial $\wkn{m}{}$ in the expression for $\Psi_j$ by its
  corresponding auxiliary Wick monomial $\wkn{m}{\delta}$ yields
  elements $\Psi_{j,\delta} \in {\cal D}_\omega$ which converge to $\Psi_j$
  for $\delta\rightarrow 0$. Now let $\epsilon >0$ be given. By
  assumption there exists some $j$ with
  \[
  \| \Psi_j - \Psi\| < \frac{\epsilon}{2}, \qquad
  \|\wkn{n}{}\! (f) ( \Psi_j - \Psi) \| < \frac{\epsilon}{2}.
  \]
  Moreover, due to the continuity property of the m-point distribution
  for the auxiliary Wick monomials
  (Proposition~\ref{Prop:limitwickaux} above), there exists a
  $\delta>0$, such that
  \[
  \| \Psi_{j,\delta} - \Psi_j\| < \frac{\epsilon}{2},\qquad
  \|\wkn{n}{}\! (f) ( \Psi_{j,\delta} - \Psi_j) \| < \frac{\epsilon}{2}.
  \]
  Combining the two previous results we get
  \[
  \| \Psi_{j,\delta} - \Psi\| \leq \epsilon \qquad
  \|\wkn{n}{}\!\! (f) (\Psi_{j,\delta} - \Psi) \| \leq \epsilon
  \]
  hence, $\{\wkn{n}{}\!\! (f) \Psi,\Psi\} \in
  \overline{\Gamma(\wkn{n}{}\! (f)\restriction_{{\cal D}_\omega})}$,
  which proves the assertion.
\end{pf}
Let us now finish the argument by showing that the linear functionals
${\cal W}^{(n_1,\cdots,n_m)}_m$ are Wightman distributions.
\begin{Lemma}[Positivity]
  The hierarchy of m-point distributions for the Wick monomials
  satisfy positivity, i.e.,
  \begin{equation}\label{eq:wickpowersPositivity}
    \begin{split}
      & \sum\idotsint \bar{f}_j(x_1,\dots,x_j)\\
      & \qquad \times {\cal W}^{(m_j,\dots,m_1,n_1,\dots,n_k)}_{j+k}
       (x_j,\dots,x_1,y_1,\dots,y_k)
       \,\, f_k(y_1,\dots,y_k)
       \geq 0,
    \end{split}
  \end{equation}
  for all finite sequences $f_0$, $f_1(x_1)$, $f_2(x_1,x_2)$, $\dots$
  of test densities. As usual the upper bar denotes complex conjugation.
\end{Lemma}
\begin{pf}
  The fields $\wkn{\otimes n}{}$ obviously are
  (hermitian) Wightman fields. Let us denote their Wightman
  M-point distributions by
  \begin{multline}\label{eq:aux_Wightmandistr}
  {\cal W}^{(\otimes n_1,\cdots,\otimes n_m)}_{M=\sum n_i}
  (x_{1_1},\dots,x_{1_{n_1}},\dots,x_{m_1},\dots,x_{m_{n_m}}) \\
  :=
  (\Omega,\wkn{\otimes n_1}{}(x_{1_1},\dots,x_{1_{n_1}})\cdots \wkn{\otimes
    n_m}{}(x_{m_1},\dots,x_{m_{n_m}}) \Omega).
  \end{multline}
  The corresponding hierarchy satisfies positivity and local
  commutativity by definition.

  Consider now a finite sequence of test densities
  $\{f_i(x_1,\dots,x_i)\}$. We may use the operators $\Delta_\epsilon$
  introduced above to map every $f_i$ into an admissible test density
  $\Delta_\epsilon \circ f_i$ for $\wkn{\otimes n}{}$. It follows that
  \begin{equation}
    \label{eq:positivity_aux_Wick}
    \begin{split}
      0 \leq & \sum \idotsint
      \overline{\sideset{^{m_1}}{_\epsilon}{\Delta}\cdots
         \sideset{^{m_j}}{_\epsilon}{\Delta} \circ f_j}
       (x_{1_1},\dots,x_{1_{m_1}},\dots,x_{j_1},\dots,x_{j_{m_j}}) \\
      & \quad  \times {\cal W}^{(\otimes m_j,\cdots,\otimes m_1,\otimes
        n_1,\cdots,\otimes n_k)}_{J+K}
      (x_{j_{n_j}},\dots,x_{1_{1}},y_{1_1},\dots,y_{k_{n_k}})\\
      & \qquad \times
      \sideset{^{n_1}}{_\epsilon}{\Delta}\cdots
         \sideset{^{n_k}}{_\epsilon}{\Delta} \circ f_k
       (y_{1_1},\dots,y_{1_{n_1}},\dots,y_{k_1},\dots,y_{k_{n_k}})
    \end{split}
  \end{equation}
  where the upper left index on $\sideset{^n}{_\epsilon}\Delta$
  indicates the image space ${\E D}'(M^n)$ of the corresponding
  operator and $J$ and $K$ abbreviate $\sum m_i$ and $\sum n_i$
  respectively. We now insert Eqn.~(\ref{eq:aux_Wightmandistr})
  together with Eqn.~(\ref{eq:aux_wick_monomial}) and use the symmetry
  of the Wick ordering prescription as well as the fact that all
  operators $\Delta_\epsilon$ are real. We obtain:
  \begin{equation*}
    \begin{split}
      (\ref{eq:positivity_aux_Wick}) = &
      \sum\idotsint \bar{f}_j(x_1,\dots,x_j)\\
      & \quad \times
      (\Omega, \wkn{m_j}{\epsilon}(x_j) \cdots
      \wkn{m_1}{\epsilon}(x_1) \wkn{n_1}{\epsilon}(y_1) \cdots
      \wkn{n_k}{\epsilon}(y_k) \Omega)\\
      & \qquad \times f_k(y_1,\dots,y_k)
    \end{split}
  \end{equation*}
  Note that by Proposition~\ref{Prop:limitwickaux} the expression
  above converges to the l.h.s. of
  Eqn.~(\ref{eq:wickpowersPositivity}) in the limit $\epsilon
  \rightarrow 0$, which finishes the proof.
\end{pf}
The locality for the m-point distributions
${\cal W}^{(n_1,\cdots,n_m)}_m$ is the result of the following
\begin{Lemma}[Locality]
  The m-point distributions ${\cal W}^{(n_1,\cdots,n_m)}_m$ obey
  locality, i.e., for $f_1$, $\dots$, $f_m \in \CinfO(M,\Omega_1)$, such
  that the supports of $f_i$ and $f_{i+1}$ are space-like separated,
  \[
  {\cal W}^{(n_1,\cdots,n_m)}_m
  (f_1\otimes \cdots f_i \otimes f_{i+1} \otimes \cdots f_m)
  =
  {\cal W}^{(n_1,\cdots,n_m)}_m
  (f_1\otimes \cdots f_{i+1} \otimes f_{i} \otimes \cdots f_m)
  \]
\end{Lemma}
\begin{pf}
  As in the proof of the previous Theorem, we prove the statement for
  the auxiliary Wick monomials first:\\
  Let $f_i$ and $f_{i+1}$ be two test densities with space-like
  separated supports. Applying the appropriate operator $\Delta$ to
  $f_i$ and $f_{i+1}$ yields test densities
  \mbox{$\sideset{^{n_i}}{}{\Delta}(f_i)$} and
  \mbox{$\sideset{^{n_{i+1}}}{}{\Delta}(f_{i+1})$} whose supports are
  space-like separated, too.  Hence one can find two space-like
  separated closed bounded regions
  $U_{\sideset{^{n_i}}{}{\Delta}(f_i)}$ and
  $U_{\sideset{^{n_{i+1}}}{}{\Delta}(f_{i+1})}$ strictly containing
  the supports of $\sideset{^{n_i}}{}{\Delta}(f_i)$ and
  $\sideset{^{n_{i+1}}}{}{\Delta}(f_{i+1})$ respectively.  This
  implies the existence of a subsequence of smooth operators
  $\sideset{^{n_i}}{_\epsilon}\Delta$,
  $\sideset{^{n_{i+1}}}{_\delta}{\Delta}$ converging to
  $\sideset{^{n_i}}{}\Delta$ and $\sideset{^{n_{i+1}}}{}\Delta$
  respectively, such that $\supp
  \sideset{^{n_i}}{_\epsilon}{\Delta}(f_i)$ and $\supp
  \sideset{^{n_{i+1}}}{_\delta}{\Delta}(f_{i+1})$ are space-like
  separated for all couples $\epsilon,\delta$. Using locality for the
  original fields $\phi$ we conclude:
  \begin{multline}
    (\Omega,\wkn{n_1}{\rho}\! (f_1)\cdots \wkn{n_i}{\epsilon}\! (f_i)
    \wkn{n_{i+1}}{\delta}\! (f_{i+1})\cdots\wkn{n_m}{\rho}\! (f_m) \Omega )\\
    =
    (\Omega,\wkn{n_1}{\rho}\! (f_1)\cdots \wkn{n_{i+1}}{\delta}\! (f_{i+1})
    \wkn{n_i}{\epsilon}\! (f_{i})\cdots\wkn{n_m}{\rho}\! (f_m) \Omega ).
  \end{multline}
  By the continuity result of
  Proposition~\ref{Prop:limitwickaux} we may pass to the limit
  $\rho,\epsilon,\delta \rightarrow 0$, obtaining
  \[
  {\cal W}^{(n_1,\cdots,n_m)}_m
  (f_1\otimes \cdots f_i \otimes f_{i+1} \otimes \cdots f_m)
  =
  {\cal W}^{(n_1,\cdots,n_m)}_m
  (f_1\otimes \cdots f_{i+1} \otimes f_{i} \otimes \cdots f_m).
  \]
\end{pf}
Combining all the results obtained and using the GNS-reconstruction
Theorem we have thus proved the following
\begin{Thm}\label{Thm:WickWightman}
  The Wick monomials of the free Klein-Gordon field on a globally
  hyperbolic spacetime with respect to any quasifree state $\omega$
  satisfying the $\mu$SC are well defined
  Wightman fields on the GNS-Hilbert space of $\omega$ with core
  ${\cal D}_\omega$ and dense invariant domain ${\cal D}$ generated by applying
  finitely many smeared Wick monomials to $\Omega$.
\end{Thm}
\begin{rem}
$\phantom{c}$\par
  \begin{itemize}
    \item  Our construction of Wick monomials generalizes by
      linearity to polynomials
      which may also contain derivatives of the
      field. This is  due to the fact that derivations do not enlarge the wave
      front set of a distribution (see the Remark in
      section~\ref{sec:prim}).
    \item It should be possible to generalize the results of this  section
      to other free field examples, e.g., the free Dirac- or
      electro-magnetic field.
  \end{itemize}
\end{rem}
It is worth noting that the GNS-vacuum $\Omega$, which was used in the
normal ordering prescription above, simultaneously yields an admissable
state for the Wick monomials in the sense of
\begin{Thm}
  Every state $\omega^{(n)}$ given by the Wightman
  distributions of the Wick monomials $\wkn{n}{}$ $(\mbox{any}\,\,\,
  n\in {\Bbb N})$ satisfies the $\mu$SC.
\end{Thm}
\begin{pf}
  The m-point distributions $\omega^{(n)}_m$ are
  ``vacuum'' expectation values of product of Wick monomials in the
  sense of Proposition~\ref{Prop:limitwickaux} by construction:
  \begin{equation}
    \label{eq:omega_n-Wick}
    \omega^{(n)}_m(x_1,\cdots,x_m)
    = \sum\begin{Sb}
       a_{i,j}\\ 1\le i<j\le m
       \end{Sb}
       \left[ \prod_{1\le i<j\le m}\,
       \dfrac{\omega_2\left(x_i, x_j\right)^{a_{i,j}}}{a_{i,j}!}\right ]
       (n!)^m
  \end{equation}
  which implies that $\WF(\omega^{(n)}_m)$ is
  contained in the following set:
  \[
  \WF(\omega^{(n)}_m) \subseteq
    \bigcup_{p\subset P} \bigoplus_{r\in p}
     \left [ \WF \left( \omega_2^r \right)\oplus
      \WF \left( \omega_2^r \right)
  \right ]
  \]
  where $p$ runs over the nonempty subsets of the set of all ordered pairs
  $P=\{(i,j), 1\le i< j\le m\}$.
  Note that we are using the same notation as in
  Eqn.~(\ref{eq:wf_omega_2_r}) and that the sum of
  two wave front sets of
  the two-point distribution over the same base points
  is stable under further sums.
  Moreover the
  two-point distribution $\omega_2$ satisfies the $\mu$SC by
  assumption and the latter is conic by definition. Hence we can
  apply verbatim the argumentation in the proof of
  Proposition~\ref{Prop:qfHadam_muSC} which completes the proof of the
  Theorem.
\end{pf}
We end this paper by stating the following straightforward result on the
energy-momen\-tum tensor for free fields on a manifold.
\begin{Cor}
  The normal ordered energy-momentum tensor $T_{\mu\nu}$ of a free
  massive scalar Klein-Gordon field on a globally hyperbolic
  manifold is a Wightman field
  on the GNS-Hilbert space for any quasifree state $\omega$ satisfying
  the $\mu$SC.
\end{Cor}
\begin{pf}
  The normal ordered energy-momentum tensor is by definition a (real) Wick
  polynomial of the corresponding free field. Hence the assertion
  follows immediately from Theorem~\ref{Thm:WickWightman} above.
\end{pf}
\section{Summary and outlook}
\label{sec:summary}
In this paper the program sketched in~\cite{koehler:95} is continued.
In particular in its first part a microlocal spectrum condition
($\mu$SC) for all m-point distributions of a quantum field on a curved
spacetime is proposed. This condition might generalize the usual
Minkowski space spectrum condition to manifolds. It is shown, that the
new $\mu$SC is compatible to the former and nontrivial examples
for physical states satisfying this new condition are presented. In
the second part of this paper arbitrary Wick monomials for free scalar
fields on globally hyperbolic spacetimes with respect to globally
Hadamard states are defined rigorously. For the proofs both the
$\mu$SC and the powerful methods of H\"ormander's microlocal analysis
are needed. In a next step they should be used to formulate causal
perturbation theory on manifolds. Work is in progress in this
direction and the authors do not expect any fundamental difficulties.

On Minkowski spacetime the study of wave front sets for more
realistic models (e.g., QED) might result in a deeper understanding of
the Wightman axioms. The free field examples considered in this paper
satisfy the stronger version of our $\mu$SC suggested in the
footnote on page~\pageref{page:strongmuSC}. We conjecture that this
property distinguishes free field theories from interacting ones. We
are confident that this conjecture  can be proved at the level of
perturbation theory.

\begin{ack}
  The authors have benefited from discussions with many members of the
  II.~Institut f\"ur Theoretische Physik at the University of Hamburg.
  R.B. gratefully acknowledges financial support of the
  Graduiertenkolleg of the University of Hamburg during his stay in
  Germany. M.K. is supported by the DFG and thanks the INFN for
  financial support and hospitality during his stay in Naples. Special
  thanks are due to M. Radzikowski for interesting discussions during
  his visit in Hamburg.
\end{ack}
%
%

\begin{thebibliography}{FSW78}

\bibitem[Bor62]{borchers:62}
H.~J. Borchers.
\newblock On the structure of the algebra of the field operators.
\newblock {\em Nuovo Cimento}, 24:214, 1962.

\bibitem[DB60]{DeWitt:60}
B.~S. DeWitt and R.~W. Brehme.
\newblock Radiation damping in a gravitational field.
\newblock {\em Ann. Phys.}, 9:220--259, 1960.

\bibitem[DH72]{Hoermander:72}
J.~J. Duistermaat and L.~H{\"o}rmander.
\newblock Fourier integral operators {II.}
\newblock {\em Acta Math.}, 128:183, 1972.

\bibitem[Dim80]{dimock:80}
J.~Dimock.
\newblock Algebras of local observables on a manifold.
\newblock {\em Comm. Math. Phys.}, 77:219--228, 1980.

\bibitem[Dim82]{dimock:82}
J.~Dimock.
\newblock {D}irac quantum fields on a manifold.
\newblock {\em Trans. Am. Math. Soc.}, 269:133--147, 1982.

\bibitem[Dim92]{dimock:92}
J.~Dimock.
\newblock Quantized electromagnetic field on a manifold.
\newblock {\em Rev. Math. Phys.}, 4:223--233, 1992.

\bibitem[Dui73]{Duistermaat:FourierIntegral}
J.~J. Duistermaat.
\newblock {\em Fourier Integral Operators}.
\newblock Courant Institute of Mathematical Sciences, New York University,
  1973.

\bibitem[FH87]{fredenhagen:87}
K.~Fredenhagen and R.~Haag.
\newblock Generally covariant quantum field theory and scaling limit.
\newblock {\em Comm. Math. Phys.}, 108:91, 1987.

\bibitem[Fre92]{fredenhagen:91}
K.~Fredenhagen.
\newblock On the general theory of quantized fields.
\newblock In K.~Schm{\"u}dgen, editor, {\em Mathematical Physics {X}}, pages
  136--152, Berlin, 1992. Springer Verlag.

\bibitem[FSW78]{fullingSweenyWald:78}
S.~A. Fulling, M.~Sweeny, and R.~M. Wald.
\newblock Singularity structure of the two-point function in quantum field
  theory in curved spacetime.
\newblock {\em Com. Math. Phys.}, 63:257--264, 1978.

\bibitem[Ful89]{Fulling:aspects_of_qft}
S.~A. Fulling.
\newblock {\em Aspects of Quantum Field Theory in Curved Space-Time}.
\newblock Cambridge University Press, Cambridge, 1989.

\bibitem[Haa92]{haag:alg}
R.~Haag.
\newblock {\em Local quantum physics: Fields, particles, algebras.}
\newblock Springer, Berlin, Germany, 1992.

\bibitem[Hep69]{hepp:renormalisation}
K.~Hepp.
\newblock {\em Th{\'e}orie de la renormalisation}.
\newblock Number~2 in Lecture Notes in Physics. Springer Verlag,
  Berlin,Heidelberg, 1969.

\bibitem[HK64]{haag:64}
R.~Haag and D.~Kastler.
\newblock An algebraic approach to quantum field theory.
\newblock {\em J. Math. Phys.}, 5:848, 1964.

\bibitem[HNS84]{haagnarnhofer:84}
R.~Haag, H.~Narnhofer, and U.~Stein.
\newblock On quantum field theory in gravitational background.
\newblock {\em Comm. Math. Phys.}, 94:219--238, 1984.

\bibitem[H{\"o}r71]{Hoermander:71}
L.~H{\"o}rmander.
\newblock Fourier integral operators {I.}
\newblock {\em Acta Math.}, 127:79, 1971.

\bibitem[H{\"o}r83]{hoermander:analysisI}
L.~H{\"o}rmander.
\newblock {\em The Analysis of Linear Partial Differential Operators~{I}}.
\newblock Springer, Berlin, 1983.

\bibitem[Jun95]{junker:95}
W.~Junker.
\newblock {\em Adiabatic vacua and {H}adamard states for scalar quantum fields
  on curved spacetimes}.
\newblock PhD thesis, University of {H}amburg, 1995.

\bibitem[K{\"o}h95]{koehler:95}
M.~K{\"o}hler.
\newblock New examples for {W}ightman fields on a manifold.
\newblock {\em Class. Quant. Grav.}, 12:1413--1427, 1995.

\bibitem[KW91]{kay:91}
B.~S. Kay and R.~M. Wald.
\newblock Theorems on the uniqueness and thermal properties of stationary,
  nonsingular, quasifree states on spacetimes with a bifurcate {K}illing
  horizon.
\newblock {\em Phys. Rep.}, 207(2):49--136, 1991.

\bibitem[Rad92]{Radzikowski:92}
M.~J. Radzikowski.
\newblock {\em The {H}adamard condition and {K}ay's conjecture in (axiomatic)
  quantum field theory on curved space-time}.
\newblock PhD thesis, Princeton University, October 1992.

\bibitem[Sat69]{Sato:69}
M.~Sato.
\newblock Hyperfunctions and partial differential equations.
\newblock In {\em Proc. Int. Conf. on Funct. Anal. and Rel. Topics}, pages
  91--94, Tokyo, 1969. Tokyo University Press.

\bibitem[Sat70]{Sato:70}
M.~Sato.
\newblock Regularity of hyperfunction solution of partial differential
  equations.
\newblock {\em Actes Congr. Int. Math. Nice}, 2:785--794, 1970.

\bibitem[Uhl62]{uhlmann:62}
A.~Uhlmann.
\newblock {\"U}ber die {D}efinition der {Q}uantenfelder nach {W}ightman und
  {H}aag.
\newblock {\em Wiss. Zeitschrift Karl Marx Univ.}, 11:213, 1962.

\bibitem[Ver94]{Verch:94}
R.~Verch.
\newblock Local definiteness, primarity and quasiequivalence of quasifree
  {H}adamard quantum states in curved spacetime.
\newblock {\em Comm. Math. Phys}, 160:507--536, 1994.

\bibitem[Wal78]{wald:78}
R.~M. Wald.
\newblock Trace anomaly of a conformally invariant quantum field in curved
  spacetime.
\newblock {\em Phys. Rev. D}, 17(6):1477--1484, 1978.

\bibitem[Wal94]{wald:QFT}
R.~M. Wald.
\newblock {\em Quantum field theory in curved spacetime and black hole
  thermodynamics}.
\newblock Chicago lectures in physics. Univ. Chicago Press, Chicago, USA, 1994.

\bibitem[WG64]{WightmanGarding:64}
A.~S. Wightman and L.~G{\aa}rding.
\newblock Fields as operator valued distributions in relativistic quantum
  theory.
\newblock {\em Ark. Fys}, 23(13), 1964.

\bibitem[Wol92]{wollenberg:92b}
M.~Wollenberg.
\newblock Scaling limits and type of local algebras over curved spacetime.
\newblock In W.~B. Arveson et~al., editors, {\em Operator algebras and
  topology}. Putman Research notes in Mathematics 270, Harlow: Longman, 1992.

\end{thebibliography}

\end{document}